\documentclass[journal]{IEEEtran}

\usepackage{graphicx}
\usepackage{epsfig} 
\usepackage{times} 
\usepackage{amsmath} 
\usepackage{amssymb}  
\usepackage{url}
\usepackage{algorithm,algorithmic}

\usepackage{color}

\usepackage{theorem}
\theorembodyfont{\rmfamily}

\newtheorem{lem}{Lemma}
\newtheorem{defin}{Definition}
\newtheorem{theorem}{Theorem}
\newtheorem{prop}{Proposition}
\newtheorem{assum}{Assumption}

\renewcommand{\IEEEQED}{\IEEEQEDopen}

\newcommand{\dg}{^\dagger}
\newcommand{\ol}[1]{\overline{#1}}
\newcommand{\hs}{& \hspace{-3mm}}
\newcommand{\tl}[1]{\tilde{#1}}
\newcommand{\dl}{\delta}

\begin{document}

\title{Parameterization of Retrofit Controllers}

\author{Hampei~Sasahara,~\IEEEmembership{Member,~IEEE,}
        Takayuki~Ishizaki,~\IEEEmembership{Member,~IEEE,}
        and~Jun-ichi~Imura,~\IEEEmembership{Senior Member,~IEEE}
\thanks{H.~Sasahara is with Division of Decision and Control Systems, KTH Royal Institute of Technology, Stockholm, SE-100 44 Sweden e-mail: hampei@kth.se.}
\thanks{T.~Ishizaki, and J.~Imura are with Graduate School
of Engineering, Tokyo Institute of Technology,
Tokyo, 152-8552 Japan e-mail: \{ishizaki, imura\}@sc.e.titech.ac.jp.}
\thanks{This work was supported by JST MIRAI Grant Number 18077648, Japan, and JSPS KAKENHI, Japan Grant Number 18K13774.}
\thanks{Manuscript received xxx xx, 20xx; revised xxx xx, 20xx.}}

\markboth{IEEE TAC technote,~Vol.~xxx, No.~xxx, xxx~20xx}%
{Shell \MakeLowercase{\textit{et al.}}: Bare Demo of IEEEtran.cls for IEEE Journals}

\maketitle

\begin{abstract}
This study investigates a parameterization of all retrofit controllers.
Retrofit control can accomplish modular design of control systems,
i.e., independent design of subcontrollers only with its corresponding subsystem model in a dynamical network system.
In the retrofit control framework, the network system to be controlled is regarded as an interconnected system composed of the subsystem of interest and an environment.
Existing studies have revealed that all retrofit controllers can be characterized as a constrained Youla parameterization under the technical assumption that the subsystem of interest is stable.
It has also been discovered that all of the retrofit controllers that belong to a particular tractable class have a distinctive internal structure composed of a rectifier and an internal controller under the additional technical assumption that the interconnection signal from the environment to the subsystem of interest is measurable or that the internal state of the subsystem of interest is measurable.
Due to the internal structure, the retrofit controller design problem under the assumptions can be reduced to a standard controller design problem to which existing controller synthesis techniques can be applied.
The aim of this paper is to extend the above results without the technical assumptions.
It is found that the existing developments can naturally be generalized through the Youla parameterization for unstable systems and an explicit description of the inverse system with the geometric control theory.
The result leads to the conclusion that retrofit controllers can readily be designed even in the general case.
\end{abstract}


%
\IEEEpeerreviewmaketitle

\section{Introduction}


Modern cyber technologies of sensing, communicating, and actuating physical systems enable us to handle highly complex and large-scale network systems~\cite{Rajkumar2012A,Ilic2010Modeling}.
A major obstacle of designing a controller for such large-scale systems stems from restriction of information structure within the controller to be designed.
Large efforts on designing structured controllers have been devoted to overcoming the challenge along the line of work on decentralized and distributed control~\cite{Sandell1978Survey,Bakule2008Decentralized,Siljak2011Decentralized}.
The substantial research facilitates structured controller design in a computationally efficient way.

The subsequent problem considered in this paper is on \emph{distributed design} of subcontrollers inside the whole structured controller.
The existing methods are built on the premise that there exists a unique controller designer.
However, in actual network systems, there are often multiple independent controller designers.
For example, a power grid is typically governed by multiple companies each of whom is responsible for managing the corresponding part of the grid.
Accordingly, each controller for frequency regulation is independently designed by each designer~\cite{Larsen1981Applying}.
While integrated controller design by a unique designer is referred to as centralized design, independent design of subcontrollers by multiple designers is referred to as distributed design~\cite{Langbort10}.

Although the notion of distributed design is indispensable for developing scalable controller design methods,
there are several difficulties to be resolved.
The primary difficulty of distributed design is that, from the perspective of a single controller designer only the model information of their corresponding subsystem is available while the model information of the other part, called environment here, is unknown.
Moreover, even if the entire model information is provided at some time instant, the environment possibly varies depending on other controller designers' action.
Several studies on distributed design can be found, such as the retrofit control approach~\cite{Ishizaki2018Retrofit,Ishizaki2019Modularity,Ishizaki2019Retrofit}.
the deadbeat control approach~\cite{Langbort10,Farokhi2013Optimal},
the integral quadratic constraint approach~\cite{Pates2017Scalable},
and the system level synthesis approach~\cite{Wang2018Separable,Anderson2019System}.

In this paper, we pay our attention to retrofit control~\cite{Ishizaki2019Retrofit},
where the network system to be controlled is regarded as an interconnected system composed of the subsystem of interest and an unknown environment.
Retrofit controllers are defined as the controllers that can guarantee internal stability of the entire network system for any possible environment on the premise that the network system to be controlled is originally stable or has been stabilized.
A characterization and a parameterization of all retrofit controllers have been derived 
in the existing research under the following technical assumptions:
\begin{itemize}
\item The subsystem of interest is stable.
\item The interconnection signal from the environment is measurable or the internal state of the subsystem of interest is measurable.
\end{itemize}
In more detail, it has been shown that all retrofit controllers can be characterized through the Youla parameterization with a linear constraint on the Youla parameter.
Furthermore, it has also been discovered that all of the retrofit controllers that belong to a particular tractable class have a distinctive internal structure composed of a so-called rectifier and an internal controller when abundant measurement is available as stated in the technical assumptions.

The objective of this paper is to generalize the results without the above technical assumptions.
First, we consider an interconnected system in which the subsystem of interest is possibly unstable.
By considering the Youla parameterization not for the subsystem of interest but for the closed-loop system composed of the subsystem of interest and a possible environment, we can generalize the above existing result.
It turns out that all retrofit controllers can be obtained through the above modified Youla parameterization with the linear constraint on the modified Youla parameter being the same as the constraint of the existing result.
Further, we reveal that all of the retrofit controllers that belong to the above tractable class have a similar structure through an explicit description of the inverse system obtained with the geometric control theory.
Owing to this fact, it is shown that the design problem of a retrofit controller in the tractable class can be reduced to a conventional stabilizing controller design problem even without the technical assumptions.


This paper is organized as follows.
Sec.~\ref{sec:pre} gives mathematical preliminaries.
In Sec.~\ref{sec:review}, we review the exiting retrofit control framework and pose the problems treated in this paper.
We provide a characterization of all retrofit controllers for network systems with possibly unstable subsystems in Sec.~\ref{sec:cha}.
In Sec.~\ref{sec:par}, we give a parameterization of all output-rectifying retrofit controllers, introduced below, under the general output feedback case.
Sec.~\ref{sec:conc} draws the conclusion.


\section{Mathematical Preliminaries}
\label{sec:pre}
In this section, we provide mathematical preliminaries necessary for the discussion in this paper.

{\it Notation:}
We denote the set of the real numbers by $\mathbb{R}$,
the set of the $n \times m$ real matrices by $\mathbb{R}^{n \times m}$,
the identity matrix by $I$,
the transpose of a matrix $M$ by $M^{\sf T}$,
the matrix where matrices $M_i$ for $i=1,\ldots,m$ are concatenated vertically by ${\rm col}(M_i)_{i=1}^m$,
the block-diagonal matrix whose diagonal blocks are composed of $M_i$ for $i=1,\ldots,m$ by ${\rm diag}(M_i)_{i=1}^m$,
the image of a matrix $M$ by ${\rm im}\, M$,
the kernel of a matrix $M$ by ${\rm ker}\, M$,
the set of real and rational $n \times m$ transfer matrices by $\mathcal{R}^{n \times m}$,
the set of proper transfer matrices in $\mathcal{R}^{n \times m}$ by $\mathcal{RP}^{n \times m}$,
and the set of stable transfer matrices in $\mathcal{RP}^{n\times m}$ by $\mathcal{RH}^{n \times m}_{\infty}$.
When the dimensions of the spaces are clear from the context, we omit the superscript.
A matrix $M \in \mathbb{R}^{n \times m}$ is said to be injective and surjective when ${\rm ker}\,M = \{0\}$ and ${\rm im}\, M = \mathbb{R}^n$, respectively.
We say that a controller $K$ stabilizes a system $G$ when the feedback system composed of $G$ and $K$ becomes internally stable, i.e., the transfer matrices from all inputs to all outputs belong to $\mathcal{RH}_{\infty}$~\cite[Chap.~5]{Zhou1996Robust}.
When a transfer matrix $G$ has a realization $C(sI-A)^{-1}B + D$, the shorthand is defined by
\[
 \left[
 \begin{array}{c|c}
 A & B\\ \hline
 C & D
 \end{array}
 \right]:=G.
\]


{\it Projection Matrix:}
A square matrix $M$ is said to be a projection matrix along ${\rm ker}\,M$ onto ${\rm im}\,M$ when $M^2=M$.
For a projection matrix $M$, there exists a matrix $P$ being injective such that $PP^{\dagger}=M$.
Conversely, for a matrix $P$ being injective and a left inverse $P\dg$, $PP^{\dagger}$ and $I-PP^{\dagger}$ become projection matrices.
We employ the notations $\bar{P}$ and $\bar{P}^{\dagger}$ to represent the matrices such that $\bar{P}\bar{P}^{\dagger} = I-PP^{\dagger}$ and $\bar{P}^{\dagger}\bar{P} = I$.
For the matrices, ${\rm im}\,P = {\rm ker}\, \bar{P}^{\dagger}$ and ${\rm im}\, \bar{P} = {\rm ker}\, P^{\dagger}$ hold.

{\it Youla Parameterization:}
A collection of stable transfer matrices $N_{\rm r}, M_{\rm r}, U_{\rm r}, V_{\rm r}, N_{\rm l}, M_{\rm l}, U_{\rm l}, V_{\rm l} \in \mathcal{RH}_{\infty}$ defines a doubly coprime factorization of $G \in \mathcal{RP}$ if $G=N_{\rm r}M_{\rm r}^{-1} = M_{\rm l}^{-1}N_{\rm l}$ and
\begin{equation}\label{eq:cop}
 \left[
 \begin{array}{cc}
 V_{\rm l} & -U_{\rm l}\\
 -N_{\rm l} & M_{\rm l}
 \end{array}
 \right]\left[
 \begin{array}{cc}
 M_{\rm r} & U_{\rm r}\\
 N_{\rm r} & V_{\rm r}
 \end{array}
 \right] = I.
\end{equation}
If $G$ is stabilizable and detectable, there always exists a doubly coprime factorization~\cite[Chap.~12]{Zhou1996Robust}.
Then the set of all controllers that stabilize $G$, denoted by $\mathcal{K}$, is given by
\begin{equation}\label{eq:Qpara}
 \mathcal{K} = \{K=(U_{\rm r}+M_{\rm r}Q)(V_{\rm r}+N_{\rm r}Q)^{-1}: Q \in \mathcal{RH}_{\infty}\}.
\end{equation}
An important special case is given when $G$ itself is stable.
In this case, a doubly coprime factorization of $G$ is given by $N_{\bullet}=G, M_{\bullet}=I, U_{\bullet}=0, V_{\bullet}=I$, where $\bullet \in \{\mathrm{l},\mathrm{r}\}$, and then the elements of $\mathcal{K}$ are given by $K= Q(I+GQ)^{-1}$.
We refer to $Q \in \mathcal{RH}_{\infty}$ as the Youla parameter of $K$ for $G$.

\section{Review of Existing Retrofit Control}
\label{sec:review}

In this section, we first review the retrofit control based on the formulation in~\cite{Ishizaki2019Modularity}.
Further, we pose the problems treated in this paper.

\subsection{Definition of Retrofit Controllers}
\begin{figure}[t]
\centering
\includegraphics[width = .85\linewidth]{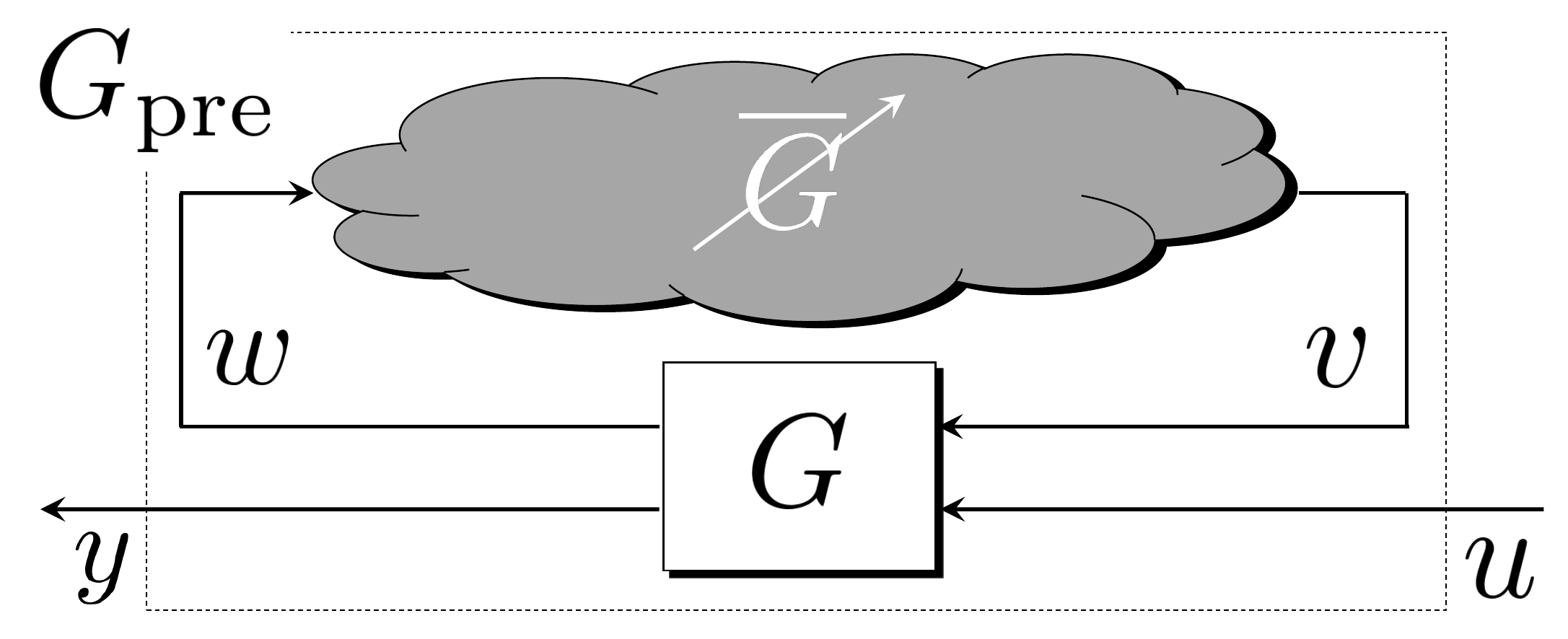}
\caption{The interconnected system composed of the subsystem of interest $G$ and the environment $\ol{G}$.
}
\label{fig:pre_sys}
\end{figure}

We consider an interconnected system in Fig.~\ref{fig:pre_sys} where
\begin{equation}\label{eq:sys_tf}
 \left[
  \begin{array}{c}
  w\\
  y
  \end{array}
 \right] = \underbrace{\left[
 \begin{array}{cc}
 G_{wv} \hs G_{wu}\\
 G_{yv} \hs G_{yu}
 \end{array}
 \right]}_{G}\left[
 \begin{array}{c}
 v\\
 u
 \end{array}
 \right]
\end{equation}
is referred to as a \emph{subsystem of interest} for retrofit control,
and
\[
 v=\ol{G}w
\]
is referred to as its \emph{environment}.
The interconnected system from $u$ to $y$ is given by
\[
 \begin{array}{rl}
 G_{\rm pre} := \hs G_{yu} + G_{yv}\ol{G}(I-G_{wv}\ol{G})^{-1}G_{wu},
 \end{array}
\]
which we refer to as a \emph{preexisting system}.
In the system representation, $w,v$ denote the interconnection signals between $G$ and $\ol{G}$ and
$y,u$ denote the measurement output and the control input.
The interconnected system can represent a large-scale network system, in which there are multiple subcontroller designers, from a single designer's viewpoint.
In this description, $G$ and $\ol{G}$ represent the subsystem corresponding to the designer and the other subsystems with interconnection of them, respectively.
For the detail of the modeling process for $G$ and $\ol{G}$, see~\cite{Ishizaki2019Modularity}.
We describe a state-space representation of the subsystem of interest~\eqref{eq:sys_tf} as
\begin{equation}\label{eq:Gss}
 G: \left\{
 \begin{array}{cl}
 \dot{x} \hs = Ax+Lv+Bu\\
 w \hs = {\it \Gamma}x\\
 y \hs = Cx
 \end{array}
 \right.
\end{equation}
where $x$ is the state of $G$.
It should be noted that, although exogenous input and evaluation output are not considered because this paper focuses just on stability analysis, our framework can also discuss control performance~\cite{Ishizaki2019Modularity}.

The purpose of retrofit control is to build a design method of the dynamical controller $K$ generating the control input according to $u=Ky$ only with the model information on the subsystem of interest $G$.
As a premise for controller design, we suppose that the preexisting system $G_{\rm pre}$ without the controller $K$ is originally stable or has been stabilized by introducing a globally coordinating controller inside the environment aside from the controller $K$ as in~\cite{Ishizaki2019Modularity}.
Under this assumption, in order to reflect the obscurity of the model information on $\ol{G}$, we introduce the set of admissible environments as
\[
 \ol{\mathcal{G}} := \{\ol{G}: G_{\rm pre}\ {\rm is\ internally\ stable.}\}.
\]
The role of the controller $K$ is to improve a control performance without losing its internal stability.
Following the discussion above, we reach the definition of retrofit controllers.
\begin{defin}
The controller $K$ is said to be a \emph{retrofit controller} if the resultant control system is internally stable for any environment $\ol{G} \in \ol{\mathcal{G}}$.
\end{defin}
The retrofit controller is a plug-and-play type controller and has an advantage that every system operator can independently introduce their own controllers.


\subsection{Characterization of Retrofit Controllers for Stable Subsystems}

A characterization of all retrofit controllers can be obtained under the assumption that the subsystem of interest $G$ is stable.
The following assumption is made.
\begin{assum}\label{assum:sta}
The subsystem of interest $G$ is stable, namely, $G \in \mathcal{RH}_{\infty}$.
\end{assum}

The first existing result indicates that all retrofit controllers can be characterized as a constrained Youla parameterization under Assumption~\ref{assum:sta}.
The following proposition holds~\cite{Ishizaki2019Modularity}.
\begin{prop}\label{prop:GQG}
Let Assumption~\ref{assum:sta} hold.
Consider the Youla parameterization of $K$ given by $K=(I+QG_{yu})^{-1}Q$ where $Q$ denotes the Youla parameter of $K$ for $G_{yu}$.
Then $K$ is a retrofit controller if and only if
\begin{equation}\label{eq:GQG}
 G_{wu}QG_{yv}=0
\end{equation}
and $Q \in \mathcal{RH}_{\infty}$.
\end{prop}

\begin{figure}[t]
\centering
\includegraphics[width = .95\linewidth]{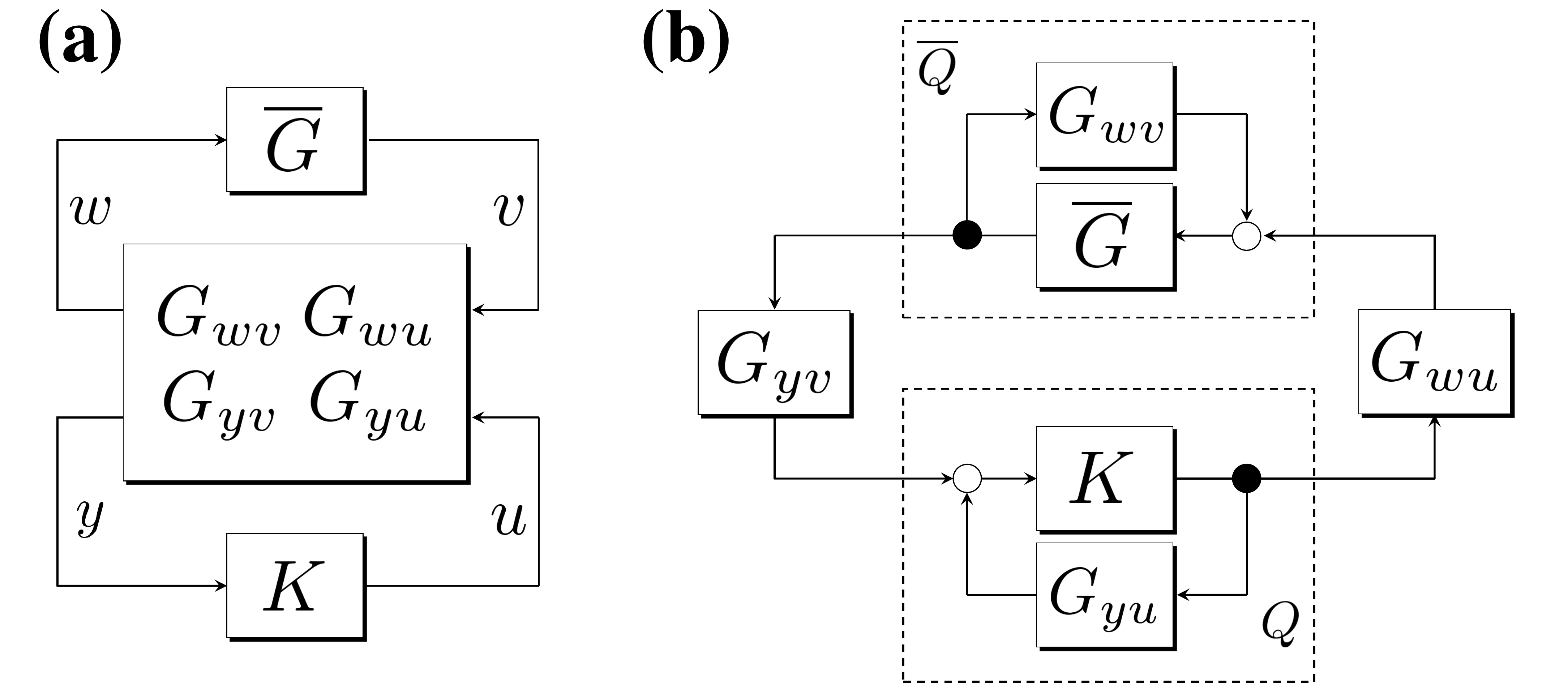}
\caption{The closed-loop systems. (a): the closed-loop system with $K$. (b): an equivalent system explicitly representing the Youla parameters $Q$ and $\ol{Q}$.}
\label{fig:cl_sys}
\end{figure}

The interpretation of the conditions is given as follows.
We first transform the closed-loop system with $K$ depicted by Fig.~\ref{fig:cl_sys}~(a) into Fig.~\ref{fig:cl_sys}~(b).
Then the transfer matrices of the subloops, depicted by boxes of broken lines in Fig.~\ref{fig:cl_sys}~(b), are exactly the same as the Youla parameters for $G_{yu}$ and $G_{wv}$ from the assumption that $G$ is stable and $G_{\rm pre}$ is internally stable.
Because $\ol{Q}$ can be taken as an arbitrary element in $\mathcal{RH}_{\infty}$ from the definition of $\ol{\mathcal{G}}$, the linear constraint~\eqref{eq:GQG} is necessary and sufficient for the internal stability.
Moreover, under~\eqref{eq:GQG}, $M_{wv}=G_{wv}$ holds, where $M_{wv}:=G_{wv}+G_{wu}(I-KG_{yu})^{-1}KG_{yv}$ denotes the closed-loop transfer matrix in terms of the interconnection signals with $K$.
This equivalence implies that the retrofit controllers can be interpreted as the controllers that keep the transfer matrix to be invariant.

Assumption~\ref{assum:sta} is essential to the claim.
Owing to the assumption, the transfer matrices with respect to the loops are equal to the Youla parameters themselves.
Without the assumption the same transformation cannot be applied for the stability analysis.

\subsection{Explicit Parameterization of Tractable Retrofit Controllers with Abundant Measurement}
\label{subsec:abun}

For retrofit controller design, it suffices to find an appropriate Youla parameter $Q \in \mathcal{RH}_{\infty}$ that satisfies the constraint~\eqref{eq:GQG} under a desired performance criterion,
but the constraint on $Q$ cannot directly be handled by a standard controller design technique.
To surmount the difficulty, a tractable class of retrofit controllers is introduced~\cite{Ishizaki2019Modularity}:
\begin{defin}\label{def:out}
The controller $K$ is said to be an \emph{output-rectifying retrofit controller} if
\begin{equation}\label{eq:KG}
 KG_{yv}=0
\end{equation}
and $Q := (I-KG_{yu})^{-1}K \in \mathcal{RH}_{\infty}$.
\end{defin}
The following proposition is obvious from the relationship between $Q$ and $K$.
\begin{prop}
Let Assumption~\ref{assum:sta} hold.
Then a controller that satisfies the conditions in Definition~\ref{def:out} is a retrofit controller.
\end{prop}

The class is tractable in the sense that all output-rectifying retrofit controllers can explicitly be parameterized with a free parameter under technical assumptions on abundance of measurement.
Two cases are considered in the existing study:
\begin{assum}\label{assum:int}
The interconnection signal $v$ is measurable in addition to the measurement output $y$.
\end{assum}
\begin{assum}\label{assum:sFB}
The measurement output $y$ is the state of $G$, namely, $x$ in~\eqref{eq:Gss}.
\end{assum}
An explicit parameterization of all output-rectifying retrofit controllers is given as follows~\cite{Ishizaki2019Modularity}.

\begin{prop}[a]
 Let Assumptions~\ref{assum:sta} and~\ref{assum:int} hold.
 Then $K$ is an output-rectifying retrofit controller if and only if there exists a proper transfer matrix $\hat{K}$ such that $K=\hat{K}R$
and $\hat{Q}:=(I-\hat{K}G_{yu})^{-1}\hat{K} \in \mathcal{RH}_{\infty}$
where $R:= [I\ -G_{yv}]$.
\end{prop}
 

\setcounter{prop}{2}

\begin{prop}[b]\label{prop:b}
 Let Assumptions~\ref{assum:sta} and~\ref{assum:sFB} hold.
 Then $K$ is an output-rectifying retrofit controller if and only if there exists a proper transfer matrix $\hat{K}$ such that $K=\hat{K}R[P^{\sf T}\ \ol{P}^{\sf T}]^{\sf T}$
and $\hat{Q}:=(I-\hat{K}\hat{G}_{yu})^{-1}\hat{K} \in \mathcal{RH}_{\infty}$
where $R := [I\ {-\hat{G}_{yv}}]$ with $\hat{G}_{yv}:= (sI-PAP^{\dagger})^{-1}PA\ol{P}\dg$ and $\hat{G}_{yu} := (sI-PAP\dg)^{-1}PB$ with a certain matrix $P$ satisfying $PL=0$.

\end{prop}

\begin{figure}[t]
\centering
\includegraphics[width = .95\linewidth]{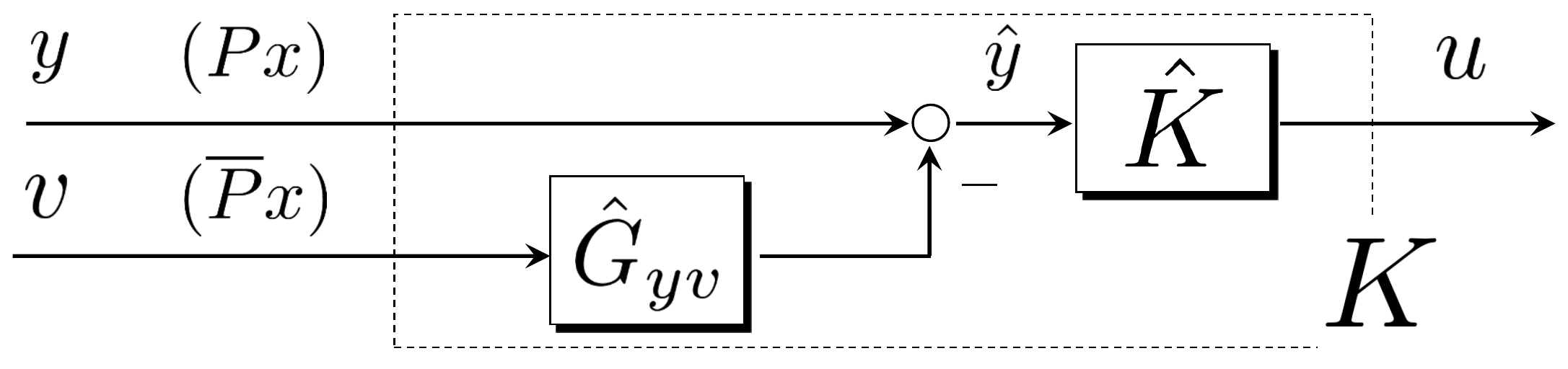}
\caption{The internal structure of all output-rectifying retrofit controllers.
In the case of measurable interconnection, $\hat{G}_{yv} = G_{yv}$ and $\hat{G}_{yu} = G_{yu}$.
In the case of state-feedback without interconnection measurement, $\hat{G}_{yv} = (sI-PAP^{\dagger})^{-1}PA\ol{P}\dg$ and $\hat{G}_{yu} = (sI-PAP\dg)^{-1}PB$.
The internal controller $\hat{K}$ is a stabilizing controller for $\hat{G}_{yu}$.
}
\label{fig:retro_str}
\end{figure}

The claims imply that, in both cases, all output-rectifying retrofit controllers have the structure illustrated by Fig.~\ref{fig:retro_str}.
The structure has two distinct characteristics.
One is that a partial measurement output $v$ ($\ol{P}x$) is used for rectifying the remaining output $y$ ($Px$) by calculating the effect from one to the other with the dynamical simulator $R=[I\ -\hat{G}_{yv}]$, which is referred to as a \emph{rectifier} because the measurement signal $y$ is rectified through the rectifier to remove the effect of $v$.
The other is that the free parameter $\hat{K}$ is characterized as a stabilizing controller for $\hat{G}_{yu}$, which is $G_{yu}$ itself or a reduced-order model of $G_{yu}$.
Since $\hat{K}$ can be an arbitrary stabilizing controller for $\hat{G}_{yu}$, the design problem of an output-rectifying retrofit controller is reduced to a standard controller design problem, which is readily handled by existing techniques.
It should be emphasized that the claims rely not only on~Assumptions~\ref{assum:int} or~\ref{assum:sFB} but also Assumption~\ref{assum:sta}.


\subsection{Objective of this Paper}\label{subsec:Q}

The objective of this paper is to generalize the existing results without the technical assumptions.
The research questions here to be addressed are as follows:
\begin{description}
\setlength{\leftskip}{-1cm}
\item{Q1:} Does there exist a simple characterization of all retrofit controllers without Assumption~\ref{assum:sta}?
\item{Q2:} What is the appropriate definition of output-rectifying retrofit controllers without Assumption~\ref{assum:sta}?
\item{Q3:} Do all output-rectifying retrofit controllers have a structure similar to Fig.~\ref{fig:retro_str} without Assumptions~\ref{assum:int} and~\ref{assum:sFB}?
\end{description}
Each of the questions corresponds to Propositions~1,~2, and~3 addressed above.
The goal of this paper is to provide solid answers to the questions.




\section{Characterization of Retrofit Controllers}
\label{sec:cha}

\begin{figure}[t]
\centering
\includegraphics[width = .95\linewidth]{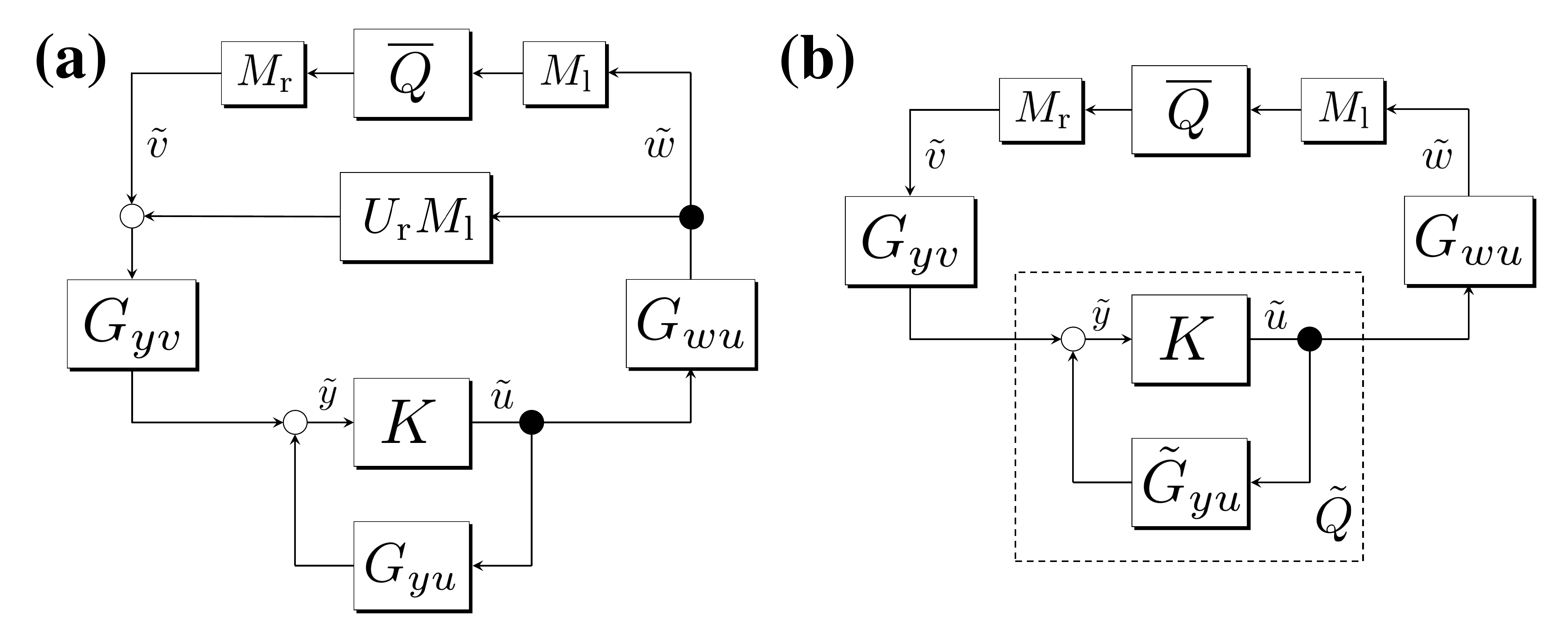}
\caption{Transformed systems from Fig.~\ref{fig:cl_sys}~(a).
(a): a block diagram where $\ol{Q}$ explicitly appears.
(b): a block diagram indicates its loop transfer matrix.
}
\label{fig:eq_sys}
\end{figure}

The aim of this section is to give an answer to Q1 by deriving a characterization of all retrofit controllers for possibly unstable subsystems.
The idea is as follows.
Consider a doubly coprime factorization of $G_{wv}$ as~\eqref{eq:cop} and denote the Youla parameter of $\ol{G}$ for $G_{wv}$ in~\eqref{eq:Qpara} by $\ol{Q}$.
Then
\[
\ol{G}(I-G_{wv}\ol{G})^{-1} = U_{\rm r}M_{\rm l} + M_{\rm r}\ol{Q}M_{\rm l} =: \mathfrak{M}(\ol{Q})
\]
holds~\cite[Sec.~4.5]{Francis1986A}.
This relationship implies that the original block diagram in Fig.~\ref{fig:cl_sys}~(a) can be transformed into that in Fig.~\ref{fig:eq_sys}~(a) where the Youla parameter $\ol{Q}$ explicitly appears.
Moreover, the transfer matrix from $\tl{v}$ to $\tl{w}$ in Fig.~\ref{fig:eq_sys}~(a), denoted by $\tl{G}_{wv}$, can be written by $\tl{G}_{wv} = G_{wu}\tl{Q}G_{yv}$ where
\begin{equation}\label{eq:Qtl}
 \tl{Q} := (I-K\tl{G}_{yu})^{-1}K,\quad \tl{G}_{yu}:= G_{yu}+G_{yv}U_{\rm r}M_{\rm l}G_{wu}.
\end{equation}
Then we obtain the further transformed block diagram in Fig.~\ref{fig:eq_sys}~(b), which explicitly represents its loop transfer matrix.
The block diagram of Fig.~\ref{fig:eq_sys}~(b) suggests that the transfer matrix from $\tl{v}$ to $\tl{w}$ must be zero for retrofit controllers as in Proposition~\ref{prop:GQG} because $\ol{Q}$ can be an arbitrary element in $\mathcal{RH}_{\infty}$.

In fact, this expectation is true.
The following theorem, our first main result, provides a necessary and sufficient condition of retrofit controllers for possibly unstable subsystems.
\begin{theorem}\label{thm:main1}
Consider a doubly coprime factorization of $G_{wv}$.
Then $K$ is a retrofit controller if and only if
\begin{equation}\label{eq:GQGgen}
 G_{wu}\tilde{Q}G_{yv} = 0
\end{equation}
and $\tl{Q} \in \mathcal{RH}_{\infty}$ where $\tl{Q}$ is defined in~\eqref{eq:Qtl}.
\end{theorem}
\begin{IEEEproof}
For stability analysis, we consider external inputs $(\dl_u,\dl_y,\dl_v,\dl_w)$ as perturbations from $(u,y,v,w)$.
The internal stability is equivalent to that the sixteen transfer matrices from $(\dl_u,\dl_y,\dl_v,\dl_w)$ to $(u,y,v,w)$ all belong to $\mathcal{RH}_{\infty}$.
Note that the transfer matrices from $(u,\dl_y,\dl_v,\dl_w)$ to $(y,v,w)$ belong to $\mathcal{RH}_{\infty}$ since the system is internally stable when $K=0$.
The transfer matrices are given by
\begin{equation}\label{eq:TFMs}
 \begin{array}{cl}
  y \hs = \{G_{yu}+G_{yv}\mathfrak{M}(\ol{Q})G_{wu}\}u + \dl_y\\
   & {}+ G_{yv}\{I+\mathfrak{M}(\ol{Q})G_{wv}\}\dl_v + G_{yv}\mathfrak{M}(\ol{Q})\dl_w,\\
  v \hs = \mathfrak{M}(\ol{Q})G_{wu}u + \mathfrak{M}(\ol{Q})\dl_w+ \{I+\mathfrak{M}(\ol{Q})G_{wv}\}\dl_v,\\
  w \hs = \{I+G_{wv}\mathfrak{M}(\ol{Q})\}G_{wu}u + \{I+G_{wv}\mathfrak{M}(\ol{Q})\}G_{wv}\dl_v\\
  & {}+ \{I+G_{wv}\mathfrak{M}(\ol{Q})\}\dl_w,
 \end{array}
\end{equation}
for the derivation of which the relationship $(I-\ol{G}G_{wv})^{-1} = I+\mathfrak{M}(\ol{Q})G_{wv}$ is used.

We here show the sufficiency.
Assume that~\eqref{eq:GQGgen} holds and $\tl{Q} \in \mathcal{RH}_{\infty}$.
For the internal stability, it suffices to show the stability of the transfer matrices in terms of $u$ because of stability of the transfer matrices in~\eqref{eq:TFMs}.
By simple algebra, we have
\begin{equation}\label{eq:TFMu}
 \begin{array}{cl}
 u \hs = \tl{Q}Xu + (I+\tl{Q}\tl{G}_{yu})\dl_u + \tl{Q}\dl_y\\
  & {}+\tl{Q}G_{yv}\{I+\mathfrak{M}(\ol{Q})G_{wv}\}\dl_v + \tl{Q}G_{wv}\mathfrak{M}(\ol{Q})\dl_w
 \end{array}
\end{equation}
with
$X:=G_{wv}M_{\rm r}\ol{Q}M_{\rm l}G_{wv}.$
Because the transfer matrices in~\eqref{eq:TFMu} except for $\tl{Q}$ appear in~\eqref{eq:TFMs},
it suffices to show $(I-\tl{Q}X)^{-1} \in \mathcal{RH}_{\infty}$ to prove the stability of the transfer matrices in~\eqref{eq:TFMu}.
Now we have $(I-\tl{Q}X)^{-1} = I+\tl{Q}X$
%
because $(\tl{Q}X)^2=0$ for any $\ol{Q}$ from~\eqref{eq:GQGgen}.
Because $G_{wv}M_{\rm r}$ and $M_{\rm l}G_{wv}$ belong to $\mathcal{RH}_{\infty}$~\cite[Theorem~4.5.1]{Francis1986A} and $\ol{Q} \in \mathcal{RH}_{\infty}$,
it turns out that $\tl{Q}X \in \mathcal{RH}_{\infty}$.
Hence $I+\tl{Q}X \in \mathcal{RH}_{\infty}$ and
$(I-\tl{Q}X)^{-1} \in \mathcal{RH}_{\infty}$.


We next show the necessity.
Assume that $K$ is a retrofit controller, i.e., the system is internally stable for any $\ol{G}\in \ol{\mathcal{G}}$.
We focus on the transfer matrix from $\dl_y$ to $u$, which is given by $T_{uy}:=(I-\tl{Q}X)^{-1}\tl{Q} \in \mathcal{RH}_{\infty}$ from~\eqref{eq:TFMu}.
First, when choosing $\ol{Q}=0$ we have $X=0$ and $T_{uy} = \tl{Q}$, which leads to that $\tl{Q} \in \mathcal{RH}_{\infty}$.
Next, $(I-\tl{Q}X)^{-1} = I+T_{uy}X \in \mathcal{RH}_{\infty}$
%
since $X \in \mathcal{RH}_{\infty}$ as shown in the sufficiency part.
From the Sylvester's determinant identity~\cite[Fact~2.14.13]{Bernstein2009Matrix} and the Nyquist stability criterion~\cite[Theorem~5.8]{Zhou1996Robust},
\[
 \begin{array}{cl}
 {\rm det}\,(I-\ol{Q}(j\omega)Y(j\omega)) \hs =  {\rm det}\, (I-\tl{Q}(j\omega)X(j\omega))\\
 \hs \neq 0, \quad {\forall}\omega \in \mathbb{R}
 \end{array}
\]
where $Y:=M_{\rm l}G_{wu}\tl{Q}G_{yv}M_{\rm r}$.
If $Y(j \omega_0)\neq 0$ for some $\omega_0 \in \mathbb{R}$, then we can construct $\ol{G} \in \ol{\mathcal{G}}$ such that ${\rm det}\,(I-\ol{Q}Y)(j\omega_0)=0$ as in the proof of the small-gain theorem~\cite[Theorem~9.1]{Zhou1996Robust}, which leads to that $(I-\tl{Q}X)^{-1} \notin \mathcal{RH}_{\infty}$ for such $\ol{G}$.
Thus $Y=0$.
This is equivalent to~\eqref{eq:GQGgen} because $M_{\rm l}$ and $M_{\rm r}$ are invertible in $\mathcal{RP}$.
\end{IEEEproof}


Under the condition~\eqref{eq:GQGgen}, $\tilde{M}_{wv}=G_{wv}$ holds, where $\tilde{M}_{wv}:=G_{wv}+G_{wu}(I-K\tilde{G}_{yu})^{-1}KG_{yv}$ represents the closed-loop transfer matrix from $v$ to $w$ with a ''stabilizing controller'' $U_{\rm r}M_{\rm l}$ inside the environment.
As in the existing result, the retrofit controllers in the general case can also be interpreted as the controllers that keep the transfer matrix among the interconnection signals with a stabilizing controller inside the environment to be invariant.
Theorem~\ref{thm:main1} provides a clear answer to Q1 posed in Sec.~\ref{subsec:Q}.
In~\eqref{eq:GQG} the variable $Q$ is simply replaced with $\tl{Q}$, the Youla parameter of $K$ for $\tl{G}_{yu}$ in~\eqref{eq:GQGgen}.
In conclusion, all retrofit controllers, even for unstable subsystems, can be characterized by the Youla parameter $\tl{Q} \in \mathcal{RH}_{\infty}$ with the linear constraint~\eqref{eq:GQGgen}.


\section{Explicit Parameterization of Output-Rectifying Retrofit Controllers}
\label{sec:par}

The aim of this section is to give answers to Q2 and Q3 in Sec.~\ref{subsec:Q} based on the result in the previous section.

\subsection{Output-Rectifying Retrofit Controllers for Unstable Subsystems}
An answer to Q2 is given as a direct conclusion of Theorem~\ref{thm:main1}.
The following theorem holds.
\begin{theorem}\label{thm:out}
If $K$ satisfies the conditions in Definition~\ref{def:out}, then $K$ is a retrofit controller for unstable subsystems as well.
\end{theorem}
\begin{IEEEproof}
Suppose that $KG_{yv}=0$ holds and $Q \in \mathcal{RH}_{\infty}$.
Then $\tl{Q}G_{yv}=(I-K\tl{G}_{yu})^{-1}KG_{yv}=0$ regardless of the choice of coprime factorization of $G_{wv}$.
Moreover, $\tl{Q} = (I-K(G_{yu}+G_{yv}U_{\rm r}M_{\rm l}G_{wu}))^{-1}K = (I-KG_{yu})^{-1}K = Q$, which implies that $\tl{Q} \in \mathcal{RH}_{\infty}$ if $Q\in \mathcal{RH}_{\infty}$.
Thus $K$ is a retrofit controller from Theorem~\ref{thm:main1}.
\end{IEEEproof}

Theorem~\ref{thm:out} implies that the existing definition of output-rectifying retrofit controllers can still be employed even when $G$ is unstable.

\subsection{Explicit Parameterization of Output-Rectifying Retrofit Controllers}

In this subsection, we give an answer to Q3.
Our aim is to investigate an intrinsic structure of all solutions of~\eqref{eq:KG} constrained by $Q\in \mathcal{RH}_{\infty}$ without Assumptions~\ref{assum:int} and \ref{assum:sFB}.
The basic idea is to apply the existing result by reproducing $v$ from a portion of $y$ through an inverse system.
Let the dimension of $y$ and $v$ be $p$ and $m$, respectively.
If $m\geq p$ and $G_{yv}$ is full-row rank, then~\eqref{eq:KG} has only the trivial solution, namely, $K=0$.
We here assume $m<p$.
Since it suffices to use $m$ independent outputs for reproducing $v$, we take the first $m$ elements of $y$ for this purpose.
Let us divide $y$ into two parts $Py$ and $\ol{P}y$ according to the matrices
\begin{equation}\label{eq:Ps}
 P:= [0\ I],\quad \ol{P}:=[I\ 0]
\end{equation}
such that
\[
 {\rm col}(y_i)_{i=1}^m = \ol{P}y
\]
where $y_i$ is the $i$th element of $y$.
By assuming that there exists an inverse system of $\ol{P}G_{yv},$ we replicate the interconnection signal $v$ according to
\[
 \hat{v}:=(\overline{P}G_{yv})^{-1} \overline{P}y.
\]
Since $\hat{y}=y-G_{yv}v$ in Fig.~\ref{fig:retro_str},
substituting $\hat{v}$ into $v$ in this equation yields
\[
 \begin{array}{cl}
 \hat{y} & \hspace{-3mm} := y-G_{yv}\hat{v} = y-G_{yv}(\overline{P}G_{yv})^{-1} \overline{P}y\\
  & \hspace{-3mm} = (P^{\dagger}P+\overline{P}^{\dagger}\overline{P})(y-G_{yv}(\overline{P}G_{yv})^{-1} \overline{P}y)\\
  & \hspace{-3mm} = P^{\dagger}\Xi y
 \end{array}
\]
with
\begin{equation}\label{eq:Xi}
 \Xi:=P(I-G_{yv}G_{yv}^{\dagger})
\end{equation}
where $G_{yv}^{\dagger}:=(\overline{P}G_{yv})^{-1} \overline{P}$ is a left inverse of $G_{yv}$.
In the following discussion, we show that all output-rectifying retrofit controllers can be constructed through the above procedure with an appropriate inverse system and also that the controller has an internal structure similar to Fig.~\ref{fig:retro_str} in the state-space representation.

First of all, we introduce the notion of relative degree for multi-input and multi-output systems.
\begin{defin}
Consider a strictly proper transfer matrix $G \in \mathcal{RP}^{p \times m}$.
Let $(A,B,C)$ be a realization of $G$ and $c_i$ be the $i$th row of $C$ for $i=1,\ldots,p$.
Then $G$ is said to have \emph{relative degree} $(r_1,\ldots, r_p)$
if for $i=1,\ldots,p$
\[
 c_i A^kB=0,\quad c_iA^{r_i-1}B\neq 0,\quad \forall k \leq r_i-2
\]
and ${\rm col}\left(c_iA^{r_i-1}B\right)_{i=1}^p$ is injective.
\end{defin}
The following assumption is made.
\begin{assum}\label{assum:rel}
The transfer matrix $G_{yv}$ has relative degree $(r_1,\ldots,r_p)$ satisfying $r_1\leq r_2 \leq\cdots\leq r_p$.
Moreover, $r_m < r_{m+1}$ and ${\rm col}\left(c_iA^{r_i-1}L\right)_{i=1}^m$ is injective.
\end{assum}
The assumption states that $G_{yv}$ has relative degree arranged in ascending order.
It is also assumed that the $m$ rows of $G_{yv}$ that have the lowest relative degrees are uniquely determined.
Although the latter assumption seems restrictive but actually generality is not lost as long as the partial transfer matrix that have relative degrees lower than or equal to $r_m$ is left invertible as shown in the following lemma.
\begin{lem}\label{lem:coo}
Assume that $G_{yv}$ has a relative degree $(r_1,\ldots,r_p)$ satisfying $r_1\leq\cdots\leq r_p$ and ${\rm col}\left(c_iA^{r_i-1}L\right)_{i=1}^m$ is injective.
Then there exists a matrix $T\in \mathbb{R}^{p \times p}$ such that $TG_{yv}$ satisfies Assumption~\ref{assum:rel}.
\end{lem}
\begin{IEEEproof}
Let $i,j$ be the maximum nonnegative integers such that $r_{m-i} = r_{m+j} = r_m$.
From the assumption, ${\rm dim}\, {\rm ker}\, {\rm col}(c_kA^{r_k-1}L)_{k=1}^{m-i-1} = i+1$.
Thus there exists a matrix $U\in \mathbb{R}^{(i+1) \times m}$ such that $ {\rm ker}\, U \cap {\rm ker}\, {\rm col}(c_kA^{r_k-1}L)_{k=1}^{m-i-1} = \{0\}$.
Let $\hat{T} \in \mathbb{R}^{(i+j+1) \times (i+j+1)}$ be a coordinate transformation matrix such that
\[
 \hat{T}{\rm col}(c_{k}A^{r_m-1}L)_{k=m-i}^{m+j} =
 \left[ U^{\sf T}\ 0
 \right]^{\sf T}.
\]
Let $T = {\rm diag}(I,\hat{T},I)$ and then $T$ satisfies the condition.
\end{IEEEproof}
Intuitively, $T$ is given by changing the coordinate of the outputs from $m_i$ to $m_j$ to satisfy that the transfer matrices with respect to the first $m$ outputs have the relative degree $r_m$ and the others have larger relative degrees.

We construct a left inverse of $G_{yv}$, namely, a possibly improper transfer matrix $G_{yv}\dg$ such that $G_{yv}\dg G_{yv}=I$.
With $\ol{P}$ in~\eqref{eq:Ps}, define $\xi$ as the derivatives of $\ol{P}y$ by
\begin{equation}\label{eq:xiDy}
 \xi = \mathcal{D}\ol{P}y
\end{equation}
with a differential operator $\mathcal{D} := {\rm diag}\left( \mathcal{D}_i\right)_{i=1}^m$ where
\[
 \mathcal{D}_i := {\rm col}\left( \dfrac{d^{j-1}}{dt^{j-1}} \right)_{j=1}^{r_i}.
\]
Then from~\eqref{eq:Gss} we have $\xi = Sx$ with
\[
 S :=
 {\rm col}
 \left(
 {\rm col}
 \left(
 c_iA^{j-1}
 \right)_{j=1}^{r_i}
 \right)_{i=1}^m.
\]
It can be shown that $S$ is surjective and there exists $\ol{S}$ such that $S$ and $\ol{S}$ complete the coordinates and $\ol{S}L=0$~\cite{Mueller2009Normal}.
Considering the coordinate transformation $x \mapsto (z,\xi)$ with $z=\ol{S}x$,
we have
\begin{equation}\label{eq:normalform}
 \ol{P}G_{yv}: \left\{
 \begin{array}{cl}
 \left[
 \begin{array}{c}
 \dot{z}\\
 \dot{\xi}
 \end{array}
 \right] \hs = \left[
 \begin{array}{cc}
 \ol{S}A\ol{S}\dg & \ol{S}AS\dg\\
 SA\ol{S}\dg & SAS\dg
 \end{array}
 \right]\left[
 \begin{array}{c}
 z\\
 \xi
 \end{array}
 \right]
 +
 \left[
 \begin{array}{c}
 0\\
 SL
 \end{array}
 \right]v\\
 \ol{P}y \hs = \ol{P}CS\dg \xi
 \end{array}
 \right.
\end{equation}
where $SL$ is injective.
The state-space representation~\eqref{eq:normalform} is referred to as a normal form~\cite{Mueller2009Normal}.
Let us suppose that the signals in~\eqref{eq:normalform} satisfy the equations.
Since $SL$ is injective, $v$ is uniquely determined by
\[
\begin{array}{cl}
 v \hs = (SL)\dg(-SA\ol{S}\dg z -SAS\dg \xi + \dot{\xi})\\
  \hs = -(SL)\dg SA\ol{S}\dg z - (SL)\dg SAS\dg \xi + (SL)\dg \dfrac{d}{dt}\mathcal{D}\ol{P}y
\end{array}
\]
from~\eqref{eq:xiDy}.
Since $(\ol{P}G_{yv})^{-1} \ol{P}G_{yv}=I$ and $\ol{(SL)}\dg SA\ol{S}\dg=0$, we have a left inverse $G_{yv}\dg:=(\ol{P}G_{yv})^{-1} \ol{P}$ by
\begin{equation}\label{eq:linv}
 G_{yv}\dg = \left[
 \begin{array}{cc|c}
 \ol{S}A\ol{S}\dg & \ol{S}AS\dg & 0\\
 0 & \ol{\tl{L}}\ol{\tl{L}}\dg SAS\dg & \tl{L}\tl{L}\dg \\ \hline
 -\tl{L}\dg SA\ol{S}\dg& - \tl{L}\dg SAS\dg & \tl{L}\dg
 \end{array}
 \right]sD_{\xi}\ol{P}
\end{equation}
where $\tl{L}:=SL$ and $D_{\xi} = {\rm diag}\left( [1\ \cdots\ s^{r_i-1}]^{\sf T} \right)_{i=1}^m$, which is given by the Laplace transformation of $\mathcal{D}$.

As a preliminary step, we prove that all solutions of~\eqref{eq:KG} in $\mathcal{RP}$ are characterized through $\Xi$ in~\eqref{eq:Xi} with the inverse system obtained above.
\begin{prop}\label{prop:1}
Under Assumption~\ref{assum:rel},
the relationship~\eqref{eq:KG}
holds if and only if there exists
a proper transfer matrix $\hat{K}$
such that
\begin{equation}\label{eq:KXi}
 K = \hat{K}\Xi.
\end{equation}
\end{prop}
For proving Proposition~\ref{prop:1}, we show the following lemma.
\begin{lem}\label{lem:1}
Let $G_{yv}\dg$ be a left inverse of $G_{yv}$ such that $G_{yv}G_{yv}\dg$ is proper.
Then~\eqref{eq:KG} holds for a proper transfer matrix $K$ if and only if there exists a proper transfer matrix $\hat{K}$ such that $K=\hat{K}\Xi_{\Pi}$ with $\Xi_{\Pi} := \Pi(I-G_{yv}G_{yv}\dg)$
where $\Pi$ is a matrix being surjective and satisfying $\Pi G_{yv}(s)G_{yv}(s)\dg|_{s=+\infty} = 0$.
\end{lem}

\begin{IEEEproof}
Let the dimensions of the signals $v,y,u$ be $m,p,q$, respectively.
As a preparation, we give all solutions to~\eqref{eq:KG} in $\mathcal{R}^{q\times m}$.
Because $\mathcal{R}^{q\times m}$ is a vector space on the field of scalar transfer functions $\mathcal{R}$ and ${\rm dim}\,({\rm im}\, G_{yv}) = m$ since its left inverse exists,
the fundamental theorem on homomorphisms leads to that all solutions in $\mathcal{R}$ can be represented by $K=\hat{K}\Xi_{\Pi}$ with $\hat{K} \in \mathcal{R}^{q\times (p-m)}$ if $\Xi_{\Pi}$ satisfies $ \Xi_{\Pi} G_{yv}=0$ and ${\rm im}\,\Xi_{\Pi} = \mathcal{R}^{p-m}$.
The former condition is obvious.
The condition on ${\rm im}\, \Xi_{\Pi}$ is also satisfied, because $\Xi_{\Pi}$ is right-invertible since the feedthrough term of $\Xi_{\Pi}$, $\Pi$ itself, is right-invertible.
Thus all solutions to~\eqref{eq:KG} in $\mathcal{R}^{q\times m}$ can be characterized by $K=\hat{K}\Xi_{\Pi}$.

We here show that the parameter $\hat{K}$ in the characterization is proper if and only if $K$ is proper.
Because $\Xi_{\Pi}$ is proper from the assumption, $K$ is proper if $\hat{K}$ is proper.
Conversely, because $\Xi_{\Pi}$ is right-invertible there exists a proper transfer matrix $\Xi_{\Pi}\dg$ such that $\Xi_{\Pi} \Xi_{\Pi}\dg=I$, $\hat{K} = K\Xi_{\Pi}\dg$ and $\hat{K}$ is proper if $K$ is proper.
Therefore, the claim holds.
\end{IEEEproof}

Lemma~\ref{lem:1} clarifies a class of a matrix $\Pi$ and an inverse system $G_{yv}^\dagger$ with which $\Pi(I-G_{yv}G_{yv}^{\dagger})$ can characterize the solution space of~\eqref{eq:KG} in $\mathcal{RP}$.
From Assumption~\ref{assum:rel}, $P$ defined in~\eqref{eq:Ps} and the inverse system in~\eqref{eq:linv} satisfy the condition required in Lemma~\ref{lem:1}, and hence Proposition~\ref{prop:1} holds.

\begin{IEEEproof}[Proof of Proposition~\ref{prop:1}]
From Lemma~\ref{lem:1}, it suffices to show that $G_{yv}G_{yv}\dg$ is proper and $PG_{yv}(s)G_{yv}(s)\dg|_{s=+\infty}=0$ since $P$ is obviously surjective.

Note that $G_{yv}G_{yv}\dg = P\dg PG_{yv}G_{yv}\dg+\ol{P}\dg \ol{P}G_{yv}G_{yv}\dg$.
From Assumption~\ref{assum:rel}, which guarantees $r_i > r_m$ for $i=m+1,\ldots,p$, $P\dg PG_{yv}G_{yv}\dg$ is strictly proper.
The inverse~\eqref{eq:linv} satisfies $G_{yv}\dg=(\ol{P}G_{yv})^{-1} \ol{P}$ and hence we have $\ol{P}\dg\ol{P}G_{yv}G_{yv}\dg=\ol{P}\dg\ol{P}$,
which is proper.
Therefore $G_{yv}G_{yv}\dg =P\dg PG_{yv}G_{yv}\dg +\ol{P}\dg\ol{P}$ is proper and $PG_{yv}(s)G_{yv}\dg (s)|_{s=+\infty} = 0$.
\end{IEEEproof}

\begin{figure}[t]
\centering
\includegraphics[width = .95\linewidth]{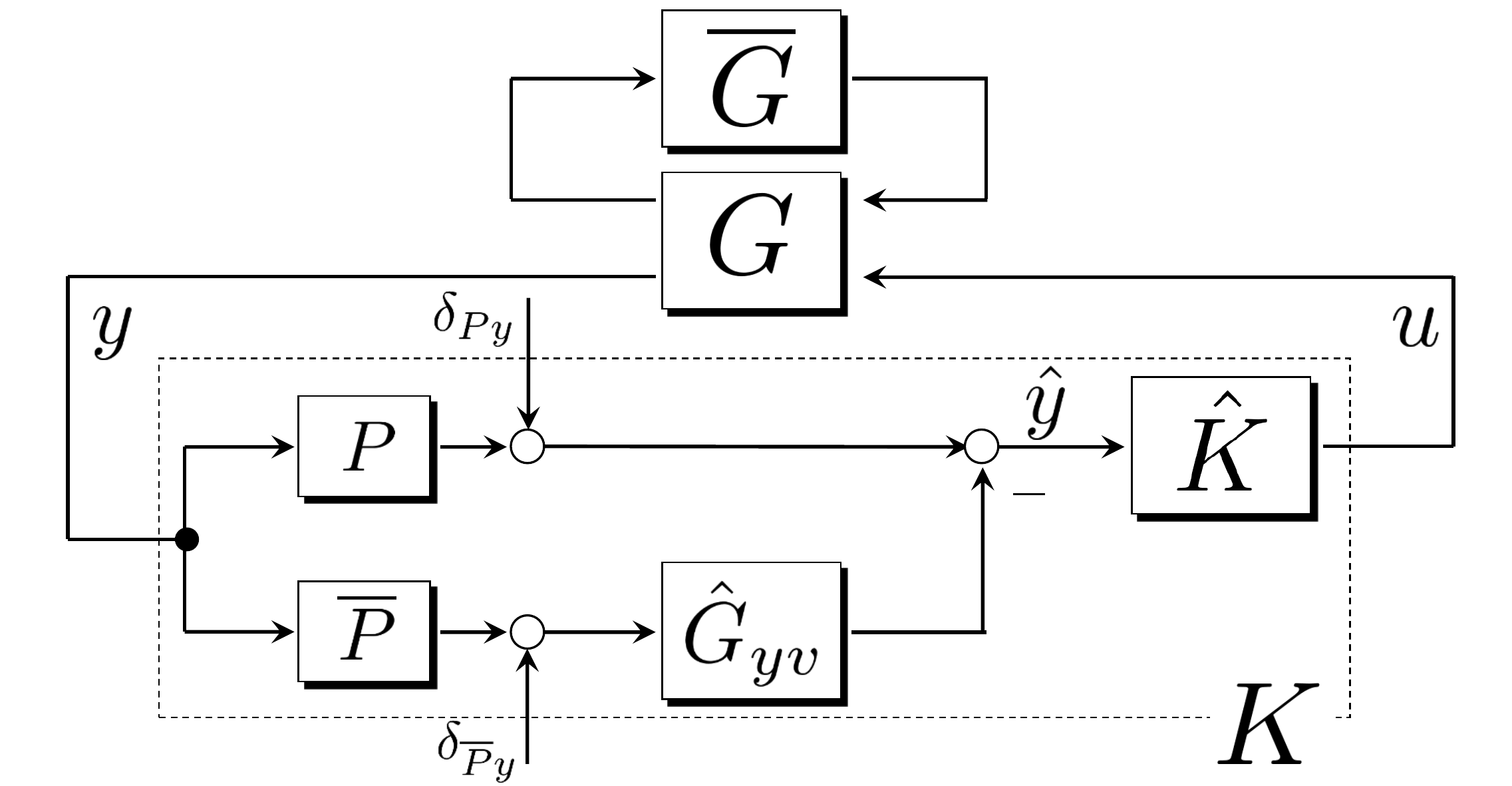}
\caption{The block-diagram of the entire closed-loop system with an output-rectifying retrofit controller, whose structure is shown in Proposition~\ref{prop:Xi}, with measurement noise.
}
\label{fig:retro_str_noise}
\end{figure}

Subsequently, we show the following proposition that unveils an internal structure of $\Xi$ in the state-space representation.
\begin{prop}\label{prop:Xi}
The relationship $\Xi=R[P^{\sf T}\ \ol{P}^{\sf T}]^{\sf T}$ holds where
\begin{equation}\label{eq:R}
 R := [I\ -\hat{G}_{yv}]
\end{equation}
with
\begin{equation}\label{eq:hatGyv}
 \begin{array}{l}
 \hat{G}_{yv} := \left[
 \begin{array}{c|c}
 \hat{A} & \ol{S}AS\dg \\ \hline
 \hat{C} & 0
 \end{array}
 \right]D_{\xi} \in \mathcal{RP},
 \end{array}
\end{equation}
$\hat{A} := \ol{S}A\ol{S}\dg,$ and $\hat{C} := PC\ol{S}\dg.$
\end{prop}
\begin{IEEEproof}[Proof of Proposition~\ref{prop:Xi}]
From Assumption~\ref{assum:rel}, $PCS\dg =0$.
Thus we have $Py = PCS\dg \xi + PC\ol{S}\dg z = PC\ol{S}\dg z$
from~\eqref{eq:normalform}.
Now because $\dot{z} = \ol{S}A\ol{S}\dg z +\ol{S}AS\dg \xi = \ol{S}A\ol{S}\dg z +\ol{S}AS\dg \mathcal{D}\ol{P}y$ from~\eqref{eq:xiDy},
it turns out that
\[
 \begin{array}{cl}
 PG_{yv} \hs = \left[
 \begin{array}{c|c}
 \hat{A} & \ol{S}AS\dg \\ \hline
 \hat{C} & 0
 \end{array}
 \right]D_{\xi} \ol{P} G_{yv} = \hat{G}_{yv} \ol{P} G_{yv}.
 \end{array}
\]
Then $PG_{yv}G_{yv}\dg =\hat{G}_{yv}\ol{P}G_{yv}(\ol{P}G_{yv})^{-1}\ol{P} = \hat{G}_{yv}\ol{P}$.
Thus $P(I-G_{yv}G_{yv}\dg) = P-\hat{G}_{yv}\ol{P}=R[P^{\sf T}\ \ol{P}^{\sf T}]^{\sf T}$.
Finally, because $G_{yv}G_{yv}\dg$ is proper, $\hat{G}_{yv} = PG_{yv}G_{yv}\dg \ol{P}\dg$ is proper as well.
\end{IEEEproof}
Proposition~\ref{prop:Xi} implies that all output-rectifying retrofit controllers have the structure illustrated in Fig.~\ref{fig:retro_str_noise}, which is the same as the existing ones shown in Fig.~\ref{fig:retro_str}.

The remaining task is to determine the class of the internal controller $\hat{K}$ such that $Q\in\mathcal{RH}_{\infty}.$
The following proposition characterizes the admissible internal controllers.
\begin{prop}\label{prop:2}
Assume that there exists a proper transfer matrix $\hat{K}$ such that~\eqref{eq:KXi} holds for a given controller $K$.
Then $Q\in \mathcal{RH}_{\infty}$ if and only if $\hat{Q} \in \mathcal{RH}_{\infty}$ and $\hat{Q}\hat{G}_{yv} \in \mathcal{RH}_{\infty}$
where
\begin{equation}\label{eq:Qhat}
 \hat{Q}:=(I-\hat{K}\hat{G}_{yu})^{-1}\hat{K}
\end{equation}
and
\[
\hat{G}_{yu} :=\left[
 \begin{array}{c|c}
 \hat{A} & \ol{S}\\ \hline
 \hat{C} & 0
 \end{array}
 \right]\left( B-AS\dg {\rm col}\left( Z_i \hat{D}_i\right)_{i=1}^m \right) \in \mathcal{RP}
\]
with
\[
 \arraycolsep=2pt
 Z_i := \left[
 \begin{array}{cccc}
 0 & \cdots & \cdots & 0\\
 c_iB & 0 &  & \vdots\\
 \vdots & \ddots & \ddots & \vdots \\
 c_iA^{r_i-2}B & \cdots & c_iB & 0
 \end{array}
 \right],\quad \hat{D}_i := {\rm col}\left(s^{j-1} I\right)_{j=1}^{r_i}.
\]
\end{prop}
\begin{IEEEproof}[Proof of Proposition~\ref{prop:2}]
We first show that $\Xi G_{yu} = \hat{G}_{y u}$.
From the definition of $\Xi$, a state-space representation of $\Xi G_{yu}$ is given by
\[
 \Xi G_{yu}:
 \left\{
 \begin{array}{cl}
 \dot{z}  \hs = \ol{S}A\ol{S}\dg z + \ol{S}AS\dg \xi + \ol{S}Bu\\
 \dot{\xi}  \hs = SA\ol{S}\dg z + SAS\dg \xi + SBu\\
 y  \hs = C\ol{S}\dg z + CS\dg \xi\\
 \dot{\zeta} \hs = \ol{S}A\ol{S}\dg \zeta + \ol{S}AS\dg \mathcal{D} \ol{P} y\\
 \hat{y} \hs = -PC\ol{S}\dg \zeta + Py
 \end{array}
 \right.
\]
with the coordinate transformation $\xi=Sx, z = \ol{S}x$.
Since $PCS\dg=0$, we have $\hat{y} = PC\ol{S}\dg(z-\zeta)$.
Let $\phi := z-\zeta$ whose dynamics is given by
\[
 \dot{\phi} = \ol{S}A\ol{S}\dg\phi + \ol{S}AS\dg(\xi-\mathcal{D}\ol{P}y)+\ol{S}Bu.
\]
Consider a state-space representation of $\ol{P}G_{yu}$ as $\dot{x}=Ax+Bu,\ \ol{P}y=\ol{P}Cx$.
Since
\[
 \dfrac{d^j y_i}{dt^j} = c_iA^jx + c_iA^{j-1}Bu + \cdots + c_iB\dfrac{d^{j-1}u}{dt^{j-1}}
\]
for $i=1,\ldots,m$ and $j=0,\ldots,r_i-1$,
it turns out that
\begin{equation}\label{eq:olPy}
\arraycolsep=1pt
 \mathcal{D}\ol{P}y = Sx + {\rm col}( Z_i \hat{\mathcal{D}}_i)_{i=1}^mu
\end{equation}
where
\[
\arraycolsep=2pt
 Z_i := \left[
 \begin{array}{cccc}
 0 & \cdots & \cdots & 0\\
 c_iB & 0 &  & \vdots\\
 \vdots & \ddots & \ddots & \vdots \\
 c_iA^{r_i-2}B & \cdots & c_iB & 0
 \end{array}
 \right],\ \hat{\mathcal{D}}_i := {\rm col}\left( \dfrac{d^{j-1}}{dt^{j-1}}I \right)_{j=1}^{r_i}.
\]
From~\eqref{eq:olPy}, we have $\dot{\phi}=\ol{S}A\ol{S}\dg\phi - \ol{S}AS\dg {\rm col}( Z_i \hat{\mathcal{D}}_i)_{i=1}^mu + \ol{S}Bu.$
Therefore, we have
\[
\arraycolsep=2pt
 \Xi G_{yu} = \left[
 \begin{array}{c|c}
 \ol{S}A\ol{S}\dg & \ol{S} \\ \hline
 PC\ol{S}\dg & 0
 \end{array}
 \right]\left( B-AS\dg {\rm col}\left( Z_i \hat{D}_i\right)_{i=1}^m \right)
\]
which is equal to $\hat{G}_{yu}$.
Moreover, because $\Xi$ and $G_{yu}$ are proper, $\hat{G}_{yu} = \Xi G_{yu}$ is proper.

Under $\Xi G_{yu} = \hat{G}_{yu}$, it turns out that $Q=\hat{Q}P-\hat{Q}\hat{G}_{yv}\ol{P} $ from $K=\hat{K}\Xi $.
Thus, if $\hat{Q}$ and $\hat{Q}\hat{G}_{yv}$ belong to $\mathcal{RH}_{\infty}$, $Q$ belongs to $\mathcal{RH}_{\infty}$.
The converse also holds because $\hat{Q}=QP\dg$ and $\hat{Q}\hat{G}_{yv} = -Q\ol{P}\dg$.
\end{IEEEproof}
Proposition~\ref{prop:2} implies that the admissible class of the internal controller $\hat{K}$ is given as controllers with which $\hat{Q}$ and $\hat{Q}G_{yv}$ are stable.
Note that those transfer matrices are obtained through $\hat{G}_{yu}$, which is a reduced-order model of $G_{yu}$.
In this sense, the internal controller $\hat{K}$ can be interpreted as a stabilizing controller for the reduced-order models of the original subsystem of interest $G$.

In summary, we obtain the following theorem, our second main result, providing a parameterization of all output-rectifying retrofit controllers.
\begin{theorem}\label{thm:main2}
The controller $K \in \mathcal{RP}$ is an output-rectifying retrofit controller if and only if there exists $\hat{K} \in \mathcal{RP}$ such that $K=\hat{K}R[P^{\sf T}\ \ol{P}^{\sf T}]^{\sf T},$ $\hat{Q}\in \mathcal{RH}_{\infty},$ and $\hat{Q}\hat{G}_{yv}\in \mathcal{RH}_{\infty}$ where $R,\hat{G}_{yv},\hat{Q}$ are defined in~\eqref{eq:R},~\eqref{eq:hatGyv},~\eqref{eq:Qhat}, respectively.
\end{theorem}
Theorem~\ref{thm:main2} gives a positive answer to Q3 in Sec.~\ref{subsec:Q}.

\subsection{Retrofit Controller Design}

Consider designing an output-rectifying retrofit controller.
Since its structure can be fixed as stated in Theorem~\ref{thm:main2},
the design problem of $K$ is reduced to the design problem of the internal controller $\hat{K}$.
However, in Theorem~\ref{thm:main2}, not only $\hat{Q}\in \mathcal{RH}_{\infty}$ but also $\hat{Q}\hat{G}_{yv} \in \mathcal{RH}_{\infty}$ are required for the class of $\hat{K}$ while only $\hat{Q}\in \mathcal{RH}_{\infty}$ is required in the existing results.
The additional condition is arisen from the absence of Assumption~\ref{assum:sta}.
The block diagram of the entire closed-loop system with an output-rectifying retrofit controller is illustrated in Fig.~\ref{fig:retro_str_noise}.
To guarantee the internal stability, it is necessary to deal with the measurement noises $\delta_{Py}$ and $\delta_{\ol{P}y}$.
Thus $\hat{Q}$ and $\hat{Q}\hat{G}_{yv}$, the closed-loop transfer matrices from $\delta_{Py}$ and $\delta_{\ol{P}y}$ to $u$, must belong to $\mathcal{RH}_{\infty}$.
In contrast, because $\hat{G}_{yv}$ can be constructed to be stable when Assumption~\ref{assum:sta} holds~\cite{Ishizaki2019Modularity}, the second condition vanishes in Propositions~\ref{prop:b} (a) and (b).

The additional condition $\hat{Q}\hat{G}_{yv} \in \mathcal{RH}_{\infty}$ seems to make the retrofit controller synthesis problem difficult because the constraint cannot be handled with a standard controller design method.
We here provide a decision policy in terms of the internal controller $\hat{K}$ without Assumption~\ref{assum:sta}.
The solution is given by simply designing a stabilizing controller for $\hat{G}_{yu}$ as shown by the following proposition.

\begin{prop}\label{col:1}
Assume that $\hat{K}$ stabilizes $\hat{G}_{yu}$.
Then both of $\hat{Q}$ and $\hat{Q}\hat{G}_{yv}$ belong to $\mathcal{RH}_{\infty}$.
\end{prop}

\begin{IEEEproof}
It is obvious that $\hat{Q}\in \mathcal{RH}_{\infty}$.
We suppose $\hat{K}$ to be a static controller for simplicity,
although the following proof can be readily extended to the case of dynamical controllers.
Let $\hat{L}$ and $\hat{B}$ be matrices such that $(\hat{A},\hat{L},\hat{C})$ and $(\hat{A},\hat{B},\hat{C})$ are realizations of $\hat{G}_{yv}$ and $\hat{G}_{yu}$, respectively.
From the assumption, $\hat{A}+\hat{B}\hat{K}\hat{C}$ is Hurwitz.
Because
\[
 \begin{array}{cl}
 \hat{Q}\hat{G}_{yv} \hs = \left[
 \begin{array}{cc|c}
 \hat{A} & 0 & \check{L}\\
 \hat{B}\hat{K}\hat{C} & \hat{A}+\hat{B}\hat{K}\hat{C} & 0\\ \hline
 \hat{K}\hat{C} & \hat{K} \hat{C} & 0
 \end{array}
 \right]\\
  \hs =
  \left[
  \begin{array}{c|c}
  \hat{A}+\hat{B}\hat{K}\hat{C} &\check{L}\\ \hline
  \hat{K}\hat{C} & 0
  \end{array}
  \right],
 \end{array}
\]
we have $\hat{Q}\hat{G}_{yv} \in \mathcal{RH}_{\infty}$.
\end{IEEEproof}

Proposition~\ref{col:1} implies that the internal stability of the closed-loop system composed of $\hat{G}_{yu}$ and $\hat{K}$ leads to the input-output stability from $\delta_{Py}$ and $\delta_{\ol{P}y}$ to $u$.
Therefore, we only have to design a stabilizing controller for $\hat{G}_{yu}$ with standard synthesis techniques for finding a suitable internal controller $\hat{K}$.


\section{Conclusion}
\label{sec:conc}
This paper has proven that existing results on retrofit control can be extended without the technical assumptions through Youla parameterization for unstable systems and an explicit description of the inverse system.
The developed results provide a retrofit controller synthesis method, to which standard controller design methods can be applied.
We have confined our attention to a subclass of retrofit controllers for controller design.
Developing a general retrofit controller design method is included in the future work, for which recent results on controller parameterization~\cite{Wang2019A,Furieri2019An,Zheng20On} would be helpful.



%





\ifCLASSOPTIONcaptionsoff
  \newpage
\fi



\bibliography{sshrrefs}
\bibliographystyle{IEEEtran}
\end{document}